# Flux free growth of superconducting FeSe single crystals

P.K. Maheshwari[1,2], L.M. Joshi[1,2], Bhasker Gahtori[1], A.K. Srivastava[1], Anurag Gupta[1]. S.P. Patnaik[3] and V.P.S. Awana[1,*]

[1]*CSIR-National Physical Laboratory, Dr. K. S. Krishnan Marg, New Delhi-110012, India*
[2]*AcSIR- Academy of Scientific & Innovative Research-NPL, New Delhi-110012, India*
[3]*School of Physical Sciences, Jawaharlal Nehru University, New Delhi-110067, India*

We report flux free growth of superconducting FeSe single crystals by an easy and versatile high temperature melt and slow cooling method for first time. The room temperature XRD on the surface of the piece of such obtained crystals showed single [101] plane of β-FeSe tetragonal phase. The bulk powder XRD, being obtained by crushing the part of crystal chunk showed majority (~87%) β-FeSe tetragonal (space group *P4/nmm*) and minority (~13%) δ-FeSe hexagonal (space group *P6$_3$/mmc*) crystalline phases. Detailed HRTEM images along with SAED (selected area electron diffraction) showed the abundance of both majority β-FeSe and minority δ-FeSe phases. Both transport (ρ-T) and magnetization (MT) exhibited superconductivity at below around 10K. Interestingly, the magnetization signal of these crystals is dominated by the magnetism of minority δ-FeSe magnetic phase, and hence the isothermal magnetization (MH) at 4K was seen to be ferromagnetic (FM) like. Transport (ρ-T) measurements under magnetic field showed superconductivity onset at below 12K, and ρ = 0 ($T_c$) at 9K. Superconducting transition temperature ($T_c$) decreases with applied field to around 6K at 7Tesla, with $dT_c/dH$ of ~0.4K/Tesla, giving rise to an $H_{c2}(0)$ value of around 50 Tesla, 30 Tesla and 20 Tesla for normal resistivity $\rho_n$ = 90%, 50% and 10% respectively, which are calculated from conventional one band Werthamer–Helfand–Hohenberg (WHH) equation. FeSe single crystal activation energy is calculated from Thermally Activated Flux Flow (TAFF) model which is found to decreases with field from 12.1meV for 0.2Tesla to 3.77meV for 7Tesla.

**\*Corresponding Author**
Dr. V. P. S. Awana, Principal Scientist, E-mail: awana@mail.npindia.org, Ph. +91-11-45609357, Fax-+91-11-45609310, Homepage awanavps.wenbs.com

# INTRODUCTION

Discovery of superconductivity in O site F doped Fe-pnictides (REFeAsO: RE = Rare Earths) [1] had been of tremendous interest to both experimental and theoretical condensed matter physicists. In fact the superconducting transition temperature ($T_c$) of Fe-pnictides to the tune of above 50K [2, 3] is second only to the famous high $T_c$ Cuprates [4, 5]. It looked like as if thunder struck again in year 2008 [1-3] after the discovery of high $T_c$ Cuprates in 1986 [4, 5]. Though a lot of research work is yet been carried out on both the above said superconducting compounds, the mechanism of superconductivity in them is yet very much elusive. This is unlike conventional superconductors including late entrant $MgB_2$ [6], which are to a large extent explainable by the electron-phonon mediated BCS (Bardeen-Cooper-Schrieffer) theory [7]. The Fe-pnictides case is even more interesting, because conventionally one feels that Fe based compounds would rather be more prone to be magnetic than superconducting. In this regards, another Fe based superconductor, namely the Fe chalcogenide i.e., FeSe entered the superconductivity kitchen in early 2009 [8]. Though the $T_c$ of FeSe and its variants viz. FeSe/Te is relatively lower to a maximum of around 20K [8-10], the same increases tremendously under moderate applied pressures to the tune of above 30K [11-13]. The $T_c$ of FeSe could in fact be increased to above 100K for $SrTiO_3$ substrate grown ultra-thin films of the same, due to high tensile stress and thus the internal chemical pressure [14,15]. The increase in $T_c$ for FeSe can also be achieved by favorable intercalation within adjacent FeSe layers to as high as above 50K [16,17].

It is clear that FeSe qualifies for its entry into the famous high $T_c$ club with its $T_c$ as high as up to 100K [14-17]. Further, its simplistic crystal structure in comparison to Fe-pnictides and high $T_c$ Cuprates calls the same to be the ideal candidate for studying the superconductivity beyond BCS [7]. In this regards, ideally the theoreticians look for the physical property data including superconductivity characterization from the single crystalline samples. This is precisely because the crystalline materials are devoid of the grain boundary related complications. In fact, the physical property data being obtained from large crystals of materials with exotic properties viz. superconductivity, thermoelectricity or photoconductivity is a feast to the theoreticians. In this regards, though the FeSe seems to be the ideal case for studying mysterious superconductivity, the growth of reasonable size crystals of the same is yet elusive. Only tiny crystals (maximum mm size) of FeSe are yet grown, with added flux (NaCl/KCl) and that also in state of art crystal growth furnaces often involving complicated heat treatments [18-21]. Infact, it is previously known that FeSe cannot be gown directly from the high temperature melt [22]. On the other hand though the added flux (NaCl/KCl) decreases the melting temperature; the foreign contamination cannot be avoided completely.

Clearly, the single crystal growth of flux free large FeSe crystals had been a challenging problem for the experimental condensed matter physicists. In this regards, a recent article reporting successful flux free growth of large FeSe crystals had been an eye opener [23]. This work got the scientific community appreciation as well [24]. Though the crystals grown were of large size, the heating schedule thus envisaged was quite complicated and the furnace employed was state of art equipment based upon traveling-solvent floating-zone (TFSZ) technique. We tried to grow flux free FeSe crystals of reasonable size from high temperature melt employing two step cooling method. Similar approach was applied recently by some of us for the growth of flux free large $FeSe_{1/2}Te_{1/2}$ superconducting crystals [25]. However, it is known previously that though large flux free crystals of $FeSe_{1/2}Te_{1/2}$ can be grown [26-29], the same is not true for FeSe [22]. After, several trials, we employed a relatively simple and easy heating schedule though with multiple steps and finally could grow flux free FeSe crystals from its melt at $1100\ ^0C$. The crystals thus grown are superconducting at below 10K. In this article, the growth and characterization of such flux free large superconducting FeSe crystals is reported. HRTEM studies along with SAED (selected area electron diffraction) clearly approved the single crystalline nature with abundance of both majority (~87%) β-FeSe and minority (~13%) δ-FeSe phases. Further studies to completely separate the majority (~87%) β-FeSe and minority (~13%) δ-FeSe phases are yet underway. Worth mentioning is the fact, that the method thus reported is checked for repeatability couple of times. We believe our timely action related to growth of flux free FeSe crystals will catch the attention of scientific community and further refined crystals could emerge as a result.

**EXPERIMENTAL**

Basically, the constituent elements Fe and Se with 4N purity are grinded in the argon filled glove box, sealed in an evacuated quartz tube, heated (rate $1^0C$/minute) in a normal automated furnace to $1100^0C$ for 24 hours with prior intermediate steps at $350^0C$ and $750^0C$ of 4hours each. This was followed by slow cooling ($5^0$C/hour) to $460^0$C and hold for 12 hours, then to $250^0$C with very slow cooling ($2^0$C/hour), subsequently the furnace is allowed to cool naturally to room temperature over a span of around 10hours. The whole process took more than 10days. The schematic flow chart of the employed heat treatment is shown in Fig. 1. X-ray diffraction (XRD) is done at room temperature using Rigaku x-ray diffractometer with CuKα radiation of 1.54184Å. The morphology of the obtained single crystal has been seen by scanning electron microscopy (SEM) images on a ZEISS-EVO MA-10 scanning electron microscope, and Energy Dispersive X-ray spectroscopy (EDAX) is employed for elemental analysis. Detailed micro-structural characterization of the FeSe crystal was carried out using a high resolution transmission electron microscope (HRTEM, model: Tecnai G2 F30 STWIN assisted with a field emission gun for the electron source at an electron accelerating voltage of 300kV). Electrical and magnetic measurements were carried out respectively on Quantum Design (QD) Magnetic Property Measurement System (MPMS) and cryogenics-PPMS down to 2K in applied fields of up to 7Tesla.

**RESULTS & DISCUSSION**

The as synthesized crystals are of few cm size (Fig. 2a). In fact when the quartz ampoule is broken, the whole material is one piece, shiny and looking to be in single crystalline form. Because the present FeSe crystals are grown from self flux method out of the stoichiometric FeSe, hence no pre-washing was required to remove the foreign flux. For various SEM (scanning electron microscope) measurements small pieces were taken from the as such obtained sample. Figure 2(b-f) show the room temperature SEM results for the studied FeSe single crystal. Figs 2 b and c show the SEM micrographs obtained from various pieces of the FeSe chunk. Clearly the slab growth of FeSe crystals can be seen in Fig. 2(b, c) micrographs. The slab like layer by layer growth persists in the studied FeSe crystal over a large area. This is similar to that as being seen recently for flux free grown $FeSe_{1/2}Te_{1/2}$ single crystals [25]. The slab like growth for FeSe crystals is known earlier as well [18-21, 23]. The compositional analysis of selected area being carried by EDX (Energy Dispersive X ray Spectroscopy) is shown in Fig. 2d. The crystal is found to be near stoichiometric i.e. close to nominal FeSe, with only a slight loss of Se. Further, Fig. 2(e, f) show near homogenous distribution of Fe and Se in the matrix. It is clear from Fig.

2(a-f) results that the presently self grown FeSe crystal is near stoichiometric with homogenous distribution of Fe, Se and the growth nature of the same is slab like.

Figure 3 exhibits detailed micro-structural characterization of FeSe single crystal being carried out using a high resolution transmission electron microscope (HRTEM, model: Tecnai G2 F30 STWIN assisted with the field emission gun for the electron source at an electron accelerating voltage of 300kV). A uniform microstructure was observed throughout in the sample under the electron beam [Fig. 3(a)]. At further higher magnifications a compact microstructure without the porosity was delineated throughout in the specimen [inset B in Fig. 3(a)]. At low magnification, a FeSe crystal of size about 16 µm in length has been displayed as inset C in Fig. 3(a). A gray-level contrast observed in the microstructure distinguishes the presence of a minor phase (δ-phase of hexagonal crystal structure of FeSe) in the matrix constituted of a t-phase of tetragonal crystal structure as a major phase (t- and δ- phases marked in Fig. 3(a)). A selected area electron diffraction pattern (SAEDP) of δ-phase along [$2\bar{1}\bar{1}0$] zone axis of a hexagonal crystal structure of FeSe (lattice parameters: a=0.37nm, c=0.58nm, space group: P6$_3$/mmc, reference: JCPDS card no. 86-2246) has been displayed as inset A in Fig. 3(a). Correspondingly, a set of important planes of δ-phase of hexagonal hkl indices: 0002, 01$\bar{1}$2, 01$\bar{1}$1, 01$\bar{1}$0 are marked as points 1,2,3,4 on the electron diffraction spots in reciprocal space in inset A of Fig. 1(a). Similar to δ-phase, a SAEDP of t-phase along [001] zone axis of a tetragonal crystal structure of FeSe (lattice parameters: a=0.38nm, c=0.55nm, space group: P4/nmm, reference: JCPDS card no. 85-0735) has been displayed as inset D in Fig. 3(b). Correspondingly, a set of important planes of t-phase of tetragonal hkl indices: 020, 110, 200 are marked as points 5,6,7 on the electron diffraction spots in reciprocal space in inset D of Fig. 3(a). Further to resolve the presence of both major tetragonal t- and minor hexagonal δ- phases in the specimen, atomic scale imaging was performed. Figure 3(b) shows a set of 200 atomic planes of tetragonal crystal structure of FeSe with the inter-planar spacing of 0.19nm, stacked at lattice scale in throughout the region. Inset E in Fig. 3(b) exhibits the inter-planar spacing of 0.29 and 0.22nm of corresponding hexagonal atomic planes of 0002 and 10$\bar{1}$2 at lattice scale.

The room temperature X-ray diffraction (XRD) patterns of the FeSe crystal being taken after powdering the same and also as such on surface of the same are shown in Fig. 4 and its inset respectively. The surface of FeSe crystal is aligned at [101] plane, see inset Fig. 4. The same orientation is seen in ref. 23 as well. The powder XRD of the studied FeSe crystal being shown in main panel Fig. 4, though mainly corresponds to the majority tetragonal (β-FeSe) phase, yet the minority hexagonal (δ-FeSe) being marked with * can also be seen. To further elucidate and quantify the presence of both phases, we carried out detailed Rietveld analysis on the powder XRD pattern of the studied FeSe crystal and the results are shown Fig. 5. The majority tetragonal phase (β-FeSe) is ~87% and the minority hexagonal (δ-FeSe) one is ~13%. The refined lattice parameters are a = b = 3.7707(3)Å and c= 5.512(3)Å for majority phase and a = 3.623(3)Å and c = 5.877(3)Å for the minority one. The rietveld refined co-ordinate positions for majority tetragonal (β-FeSe) phase are Fe (2a) at 3/4, 1/4, 0 and Se (2c) at 1/4, 1/4, 0.2594(2). The same for minority hexagonal (δ-FeSe) are Fe (2a) at 0, 0, 0 and Se (2c) at 1/3, 2/3, 1/4. The schematic unit cells for both the β-FeSe and δ-FeSe phase are shown in inset of Fig. 4. Further, the rietveld refined data for both the phases are shown in Table 1. Interestingly, the abundance of both the β-FeSe and δ-FeSe phases is seen in our HRTEM results as well, see details in Fig. 3. Clearly, the presently grown single crystals though are majority tetragonal (β-FeSe) phase, yet the minority hexagonal (δ-FeSe) is also seen embedded. To our surprise, the only other report available on flux free large FeSe single crystals [23] did not elaborate on this point, though the minority hexagonal (δ-FeSe) is seen in their powder XRD as well. Also both majority tetragonal (β-FeSe) and the minority hexagonal (δ-FeSe) phases were seen even in flux (NaCl/KCl) grown FeSe crystals as well [12]. The only distinct way to separate the β-FeSe and δ-FeSe phases is to carry out the DC magnetization measurements, because the former is superconducting (diamagnetic) and later is known to be magnetic.

The zero-field-cooled (ZFC) and field-cooled (FC) DC magnetic susceptibility of FeSe crystal at 10Oe applied field in temperature range to 2-15K is shown in Fig.6. Though, clear branching of ZFC and FC is seen below 10K, the moment is yet +ve. This is puzzling, because ideally if the studied crystal is superconducting, the same must have inferred with −ve moment as sign of the diamagnetism. Interestingly, the observed result (Fig.6) is possible if +ve magnetic background could be riding on the diamagnetic signal from superconductivity. This superimposition is clear as the ZFC moment becomes −ve below around 5K due to increasing superconducting volume fraction below $T_c$. In present case, where both majority (~87%) superconducting (β-FeSe) and minority (~13%) magnetic (δ-FeSe) are present (XRD and HRTEM results), the magnetization outcome in Fig. 6 is not surprising. In fact, in ref. 23 as well, the DC moment outcome on similar flux free grown FeSe crystals would have been the same, but unfortunately in that article only the AC susceptibility measurements (999Hz, 1Oe) are shown. The AC susceptibility measurements (333Hz, 10Oe) on present FeSe crystal are shown in inset of Fig. 6. These measurements clearly demonstrate that our FeSe crystal is clearly superconducting at below 9K. To further elucidate upon this point, we also carried out the isothermal magnetization (MH) measurements well below the superconducting transition temperature i.e., at 4K and the results are shown in Fig. 7. Clearly a ferromagnetic (FM) loop is seen with saturation moment above 1Tesla with negligible coercively of few Oe only. The expanded part of MH at near origin is shown in inset of Fig.7, indicating clearly dominating diamagnetic part for low fields (below 70Oe) and for higher fields the moment becomes +ve due to overriding positive moment from minority magnetic ($\delta$-FeSe) phase. The lower critical field ($H_{c1}$) of the studied crystal seems to be around 25Oe, which is being marked in inset of Fig.7.

To ascertain, if the ZFC and FC branching at 10K arises out of the superconducting transition, we carried out the resistivity versus temperature ($\rho$-T) measurements on FeSe crystal and the results are shown in Fig. 8. The superconductivity onset is seen at 12K and $\rho$=0 is obtained at around 9K. The studied FeSe crystal is superconducting at below 9K and the normal state conductivity is of metallic nature. The magneto transport measurements i.e., $\rho$(T)H in superconducting transition region (4K-15K) are shown in Fig. 9. The $T_c(\rho=0)$ is decreased monotonically from 9K to around 5.8K under applied field of 7Tesla. The $dT_c/dH$ is ~0.4K/Tesla. The zero temperature upper critical field $H_{c2}(0)$ is calculated applying the conventional one-band Werthamer-Helfand-Hohenberg (WHH) equation, i.e., $H_{c2}(0) = -0.693 T_c (dH_{c2}/dT)_{T=T_c}$. The calculated $H_{c2}(0)$ with different criterion of $\rho$ = 10%, 50% and 90% of the normal state resistivity is around 20Tesla, 30Tesla and 50Tesla respectively. The WHH plots are shown in Fig.10. Importantly, the $H_{c2}(0)$ values are outside the Pauli Paramagnetic limit of $1.84T_c$. This indicates that the studied crystals are heavily pinned possibly due to the presence of minority (~13%) magnetic ($\delta$-FeSe) phase along with majority (~87%) superconducting ($\beta$-FeSe) phase.

To further elucidate upon the $\rho$(T)H behaviour of FeSe crystals the thermally activated flux flow (TAFF) plots (ln$\rho$ vs 1/T) at various fields for studied FeSe crystal are shown in Fig. 11. According to Thermally Activated Flux Flow (TAFF) theory [30-31], the Arrhenius relation is given by the equation [32] $\ln\rho(T,H) = \ln\rho_0(H) - U_0(H)/k_B T$, where $\ln\rho_0(H)$ is the temperature independent constant, $k_B$ is the Boltzmann's constant and $U_0(H)$ is TAFF activation energy. The ln$\rho$ vs 1/T plot with field in TAFF region remains linear. This linear region with fields is fitted very well and is shown in Fig11. The extrapolated fitted lines intercept at same temperature i.e., at bulk $T_c$ of the crystal, which is around 11.62K. With an increase in the magnetic field, the resistivity broadening takes place due to thermally assisted flux motion. [32]. As far as Fe based superconductors are concerned, the Ba-122 compound shows least broadening due to lower thermal fluctuations, while iron pnictides i.e. $ReO_{1-x}F_xFeAs$ (Re- 1111) show wider resistivity broadening similar to that as for Cuprate YBCO compounds with an increases in field. [32-34] Excitingly $FeSe_{1-x}Te_x$ based superconductors show intermediate resistivity broadening with increases in magnetic field. [35]. The present case of self flux grown FeSe crystals seems to be closer to Ba-122 [32].

The activation energy is calculated for the different magnetic fields in the range of 0.2Tesla to 7Tesla. The variation of activation energy with field is very wide i.e., from 12.1meV for 0.2Tesla to 3.77meV for 7Tesla, this shows how the magnetic field affect the creep of thermally activated vortices. Interestingly, this activation energy is far below than the activation energy of $FeSe_{0.5}Te_{0.5}$ single crystals [25], suggesting single vortex pinning for FeSe single crystalline compound. With increasing magnetic field, the thermally activation energy follows power law i.e. $U_0 = K \times H^{-\alpha}$ where $U_0$ is field dependent. Value of α vary with field i.e., lower for lower fields and higher for higher fields, see Fig 12. $U_0$ is calculated for different fields with α = 0.21 for lower field i.e., up to 2Tesla and α= 0.62 for higher field i.e. from 3Tesla to 7Tesla. In lower fields, the weak power low decreases of $U_0(H)$ denotes single vortex pinning [36,37]. The TAFF behaviour of FeSe seems closer to Ba-122 Fe based superconductor than the Fe-pnictide and HTSc Cuprates ones.

**CONCLUSIONS**

We reported flux free growth of FeSe single crystal without any complicated heating schedule in a simple automated furnace for first time. The XRD result of FeSe single crystal shows that the crystal growth is in [101] plane. Powder XRD result showed both majority (~87%) β-FeSe tetragonal and minority (~13%) δ-FeSe hexagonal crystalline phases. HRTEM results though suggests towards the single crystalline nature but with presence of two phases. The superconductivity at around 10K is confirmed by the both MT and ρ-T measurements. $H_{c2}(0)$ value, which is calculated from conventional one-band WHH equation comes around 50Tesla for 90% of $ρ_n$ criterion. Activation energy is estimated up to 7Tesla magnetic fields with the help of TAFF model, which showed that single vortex pinning dominants in low field region. Separating the two phases i.e., (~87%) β-FeSe tetragonal and (~13%) δ-FeSe hexagonal is the priority task at hand. In an earlier study [38], the two phases were separated from flux (KCl-AlCl$_3$) grown melt of FeSe. The flux free growth of FeSe from direct melt has yet been elusive, only scant reports with specialized techniques like CVT (chemical vapour transport) on AlCl$_3$/KCl eutectic melt do exist. In this regards, the present study is an eye opener.

**ACKNOWLEDGEMENT**

Authors would like to thank their Director NPL India for his keen interest in the present work. This work is financially supported by DAE-SRC outstanding investigator award scheme on search for new superconductors. P. K. Maheshwari and L.M. Joshi thank CSIR, India for research fellowship and AcSIR-NPL for Ph.D. registration.

**Table 1: FeSe Single Crystal Rietveld Analysis**

|  | **Tetragonal Phase (Majority Phase)** | **Hexagonal Phase (Minority Phase)** |
|---|---|---|
| **Fraction %** | ~ 87 | ~ 13 |
| **Space group** | P4/nmm | P6$_3$/mmc |
| **a=b (Å)** | 3.7707(3) | 3.623(3) |
| **c (Å)** | 5.512(3) | 5.877(3) |
| **V (Å$^3$)** | 78.377 (2) | 66.815(2) |
| **Fe (2a)** | (3/4, 1/4 ,0) | (0, 0, 0) |
| **Se (2c)** | (1/4 , 1/4 , 0.2594(2)) | (1/3 , 2/3, 1/4) |
| **JCPDS card No.** | 85-0735 | 86- 2246 |

**FIGURE CAPTIONS**

**Figure 1:** Flow chart of FeSe crystal synthesis process through self flux method.

**Figure 2:** (a) Photograph of FeSe single crystals (b-c) SEM images of FeSe single crystal at 2μm and 10μm magnification (d-f) EDX quantitative analysis graph of the FeSe single crystal.

**Figure 3:** HRTEM micrographs of FeSe single crystal showing (a) bright field electron microgrpah, and (b) atomic scale image from tetragonal t-phase. Insets: (A) SAEDP along $[2\bar{1}\bar{1}0]$ zone axis of a hexagonal crystal structure, (B) high magnification micrograph, (C) a large size crystal, (D) SAEDP along [001] zone axis of a tetragonal crystal structure, and (E) atomic scale image from hexagonal δ-phase.

**Figure 4:** Powder XRD patterns of crushed powder FeSe single crystal room temperature. *Inset* view is room temperature XRD pattern of FeSe single crystal.

**Figure 5:** The room temperature observed and Rietveld fitted XRD patterns of crushed powder of FeSe single crystal. Inset view is unit cell of tetragonal and hexagonal structure.

**Figure 6:** DC magnetization plots i.e. both ZFC and FC of FeSe single crystal measured in the applied magnetic field at H = 10 Oe, inset shows AC susceptibility measurements (333Hz, 10Oe).

**Figure 7:** Isothermal MH curve with magnetic field -2 Tesla to +2 Tesla at 4 K of FeSe single crystal. *Inset* zoomed view of MH curve with magnetic field 0 Oe to 100 Oe.

**Figure 8:** Temperature dependent electrical resistivity from temperature range 300K to 5K of FeSe single crystal.

**Figure 9:** Electrical resistivity with temperature under various magnetic fields up to 7 Tesla in temperature range of 15K to 4K for FeSe single crystal.

**Figure 10:** Upper critical field ($H_{c2}$) calculated from ρ(T,H) data with 90%, 50% and 10% $ρ_n$ criteria of FeSe single crystal.

**Figure 11:** lnρ vs 1/T for different magnetic fields of FeSe single crystal, corresponding solid line are fitting of Arrhenius relation.

**Figure 12:** Field dependent of Activation energy $U_o(H)$ with solid lines fitting as power law of $U_o(H) \sim H^{-α}$.

Fig. 1

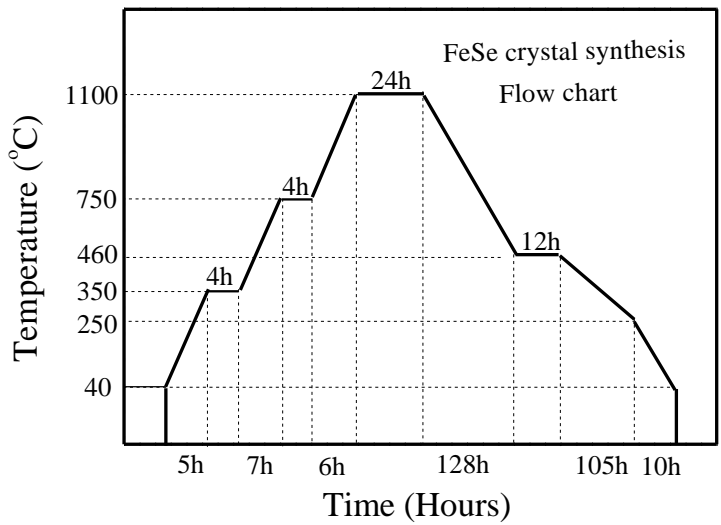

Fig. 2

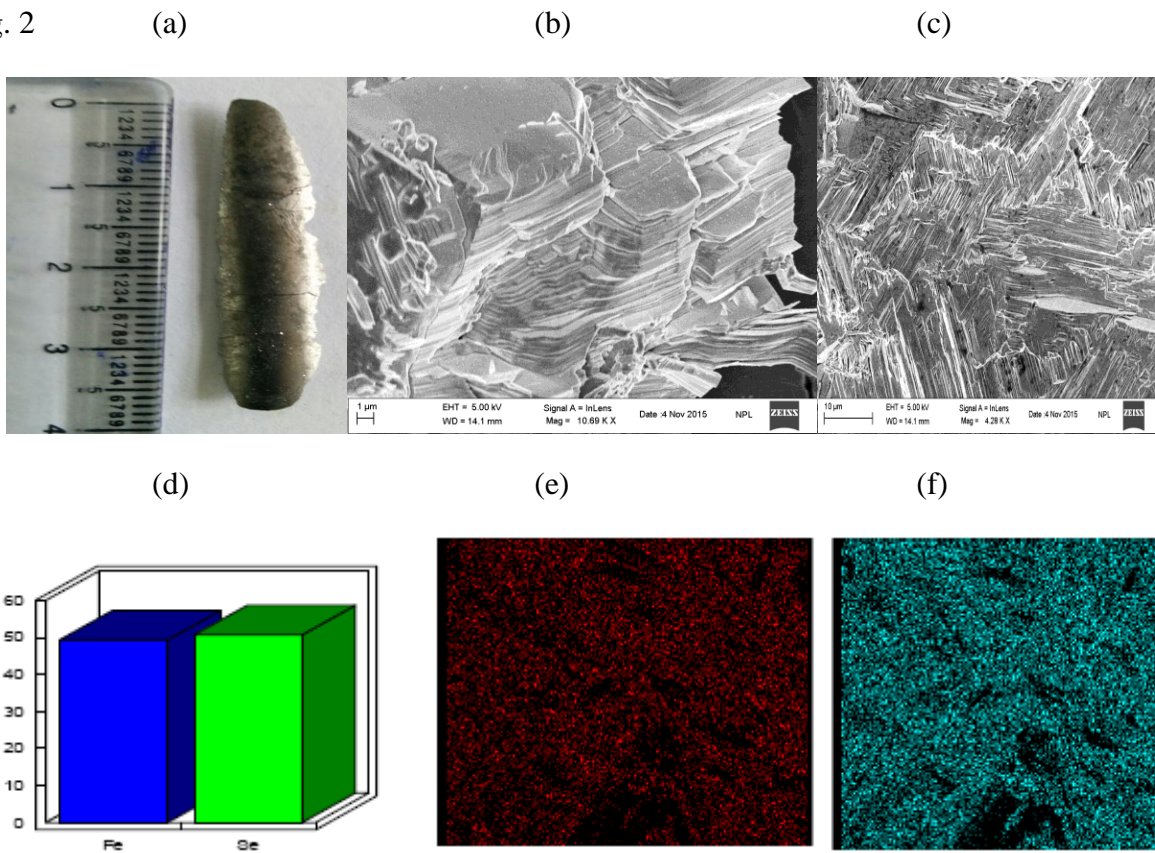

Figure 3

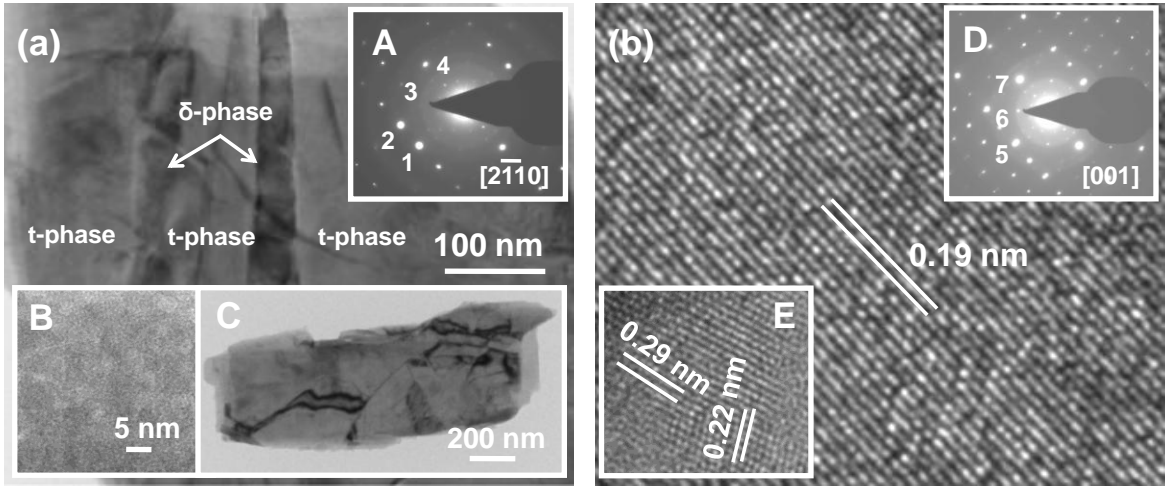

Fig. 4

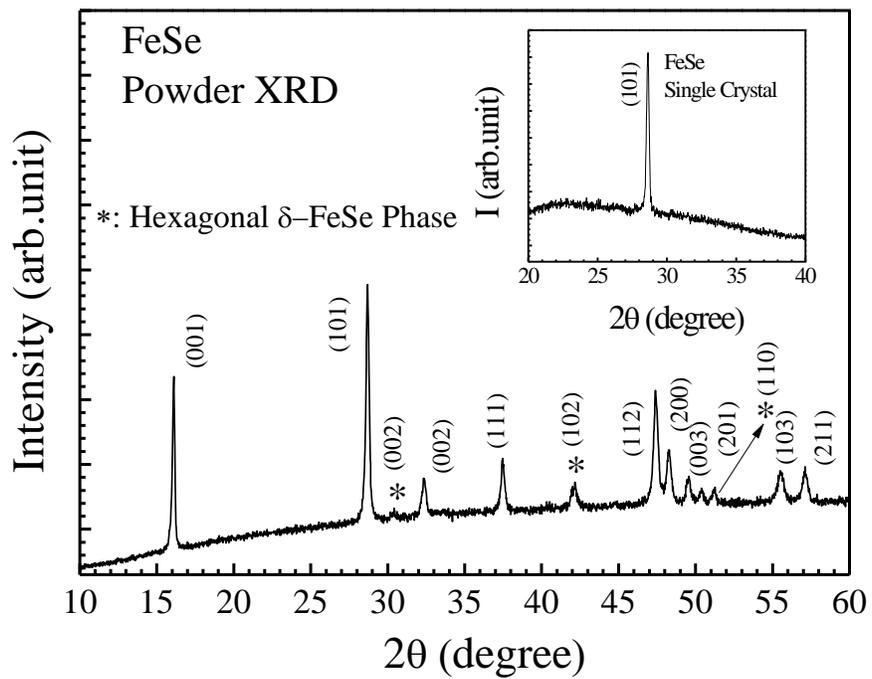

Fig. 5

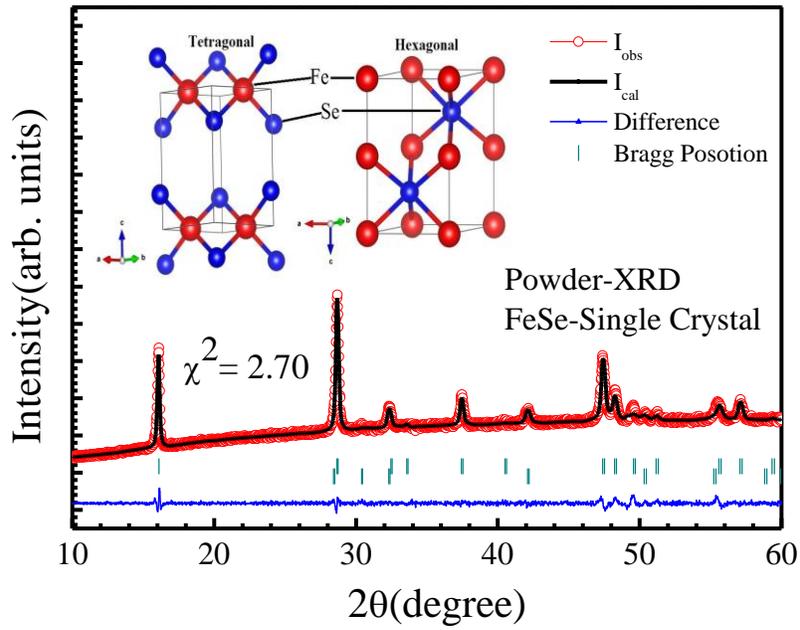

Fig. 6

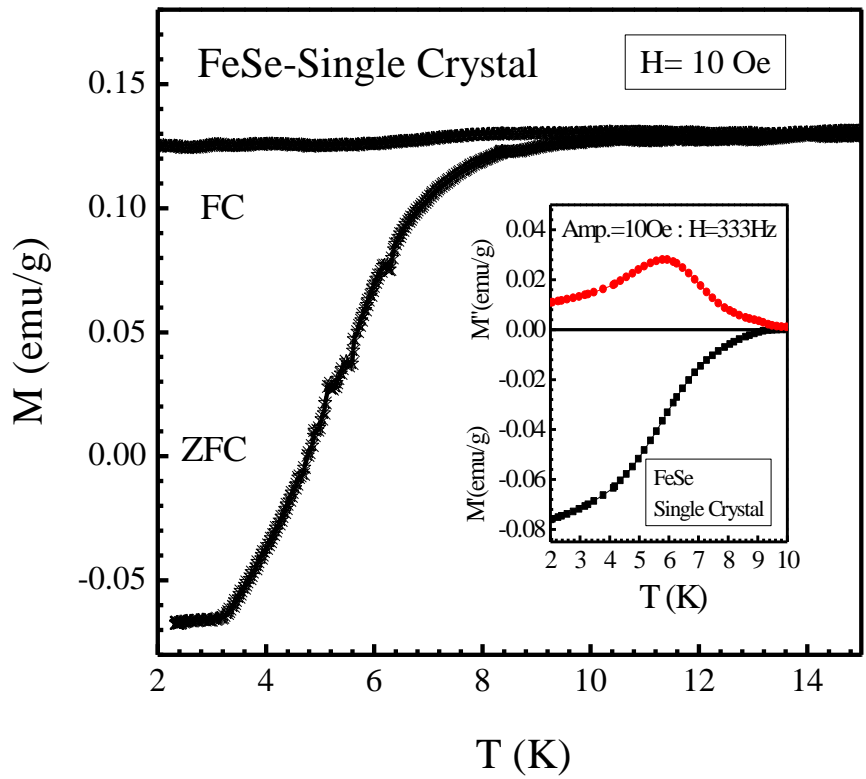

Fig. 7

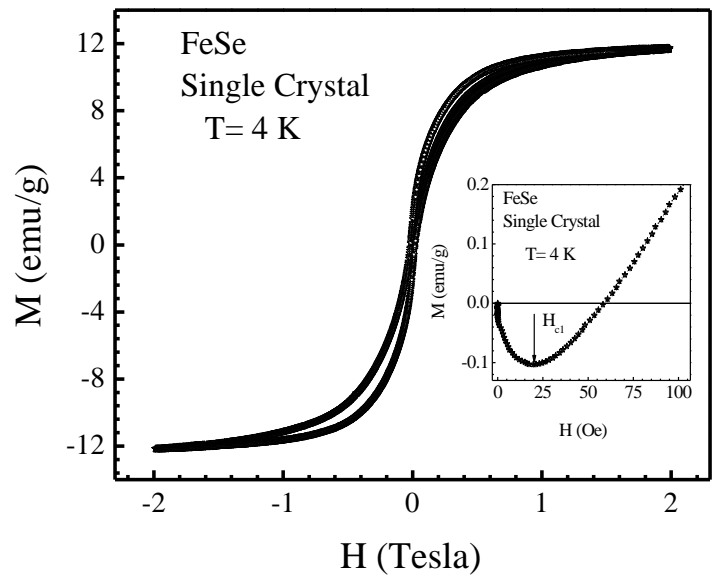

Fig. 8

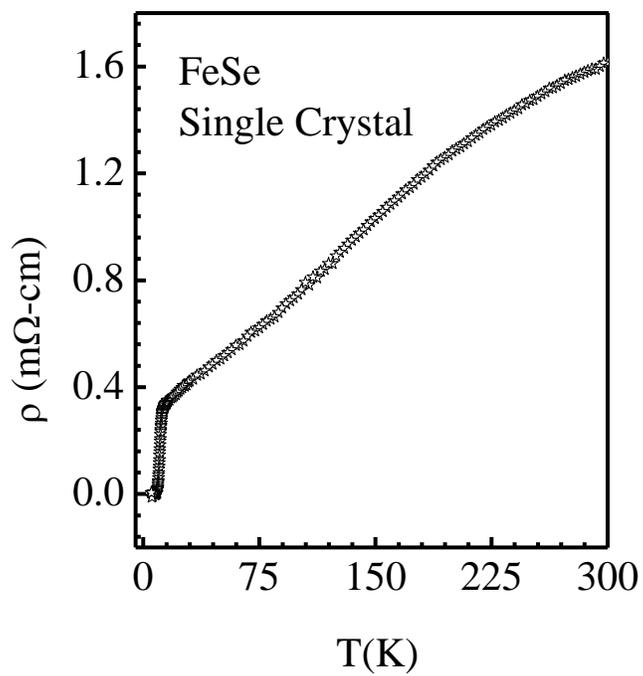

Fig. 9

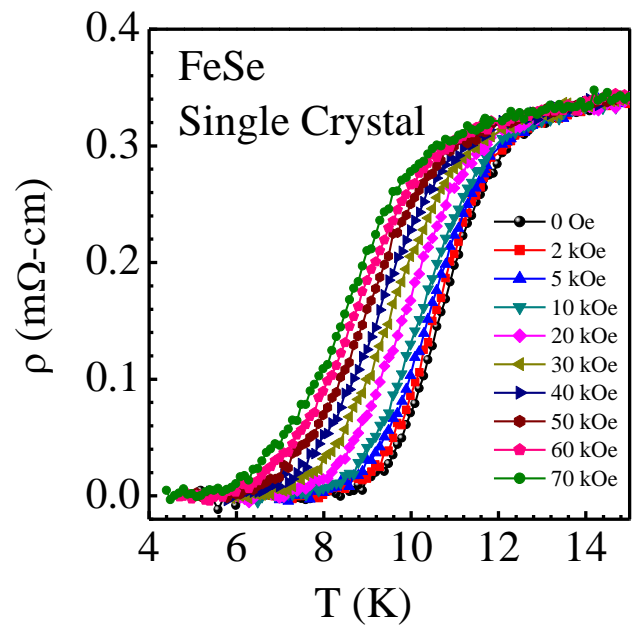

Fig. 10

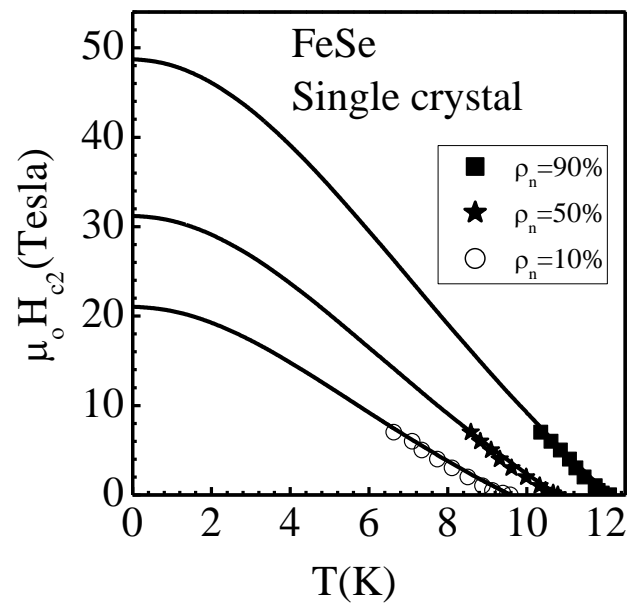

Fig.11

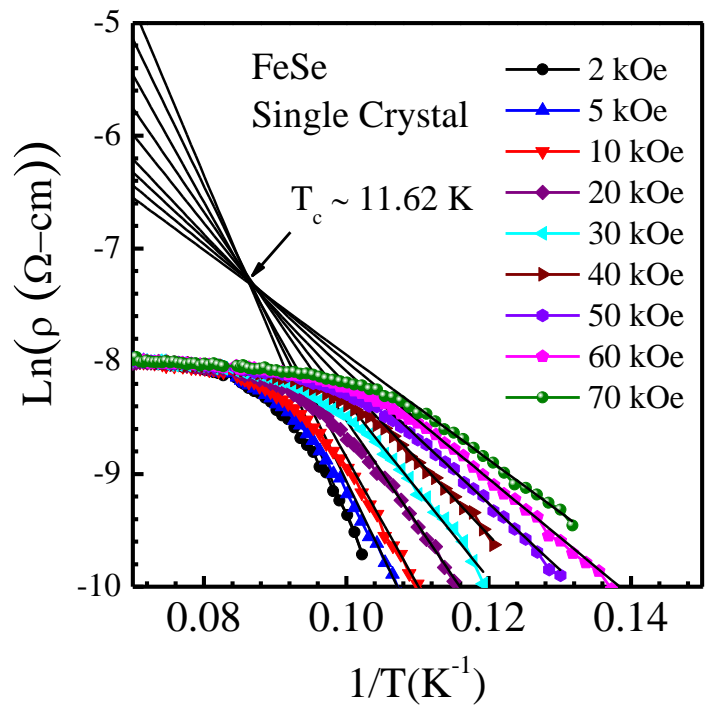

Fig. 12

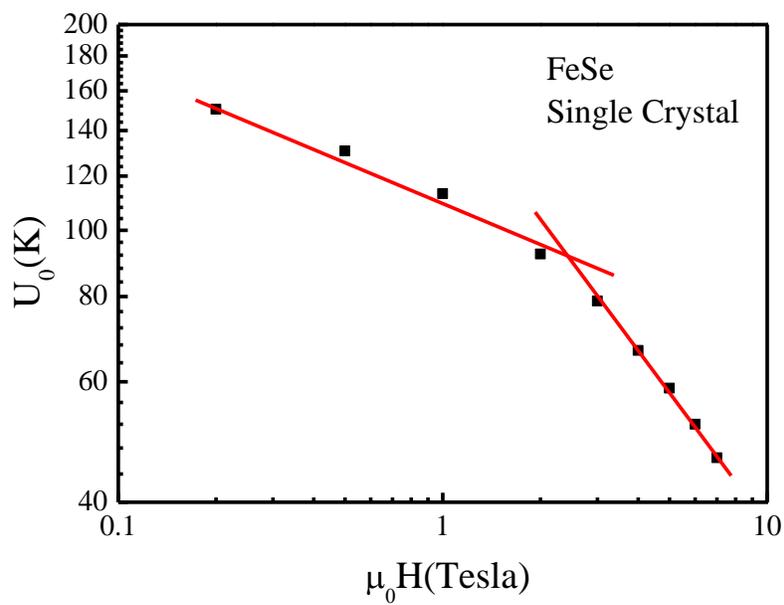